# Bianchi Models with Chaplygin Gas

G. Berköz and Ö. Sevinç

*Department of Physics, Science Faculty, Istanbul University, Vezneciler, Istanbul, Turkey*

**Abstract.** Einstein Gravitational Field Equations (EFE) of Chaplygin gas dominated Bianchi-type models are obtained by using metric approximation. The solutions of equations for a special case, namely Bianchi I model which is a generalization of isotropic Friedmann-Robertson-Walker (FRW) cosmology, are obtained. The early and late behaviours of some kinematic parameters in model are presented in graphically.

## 1. INTRODUCTION

Recent astronomical observations of type Ia supernovae, CMBR, gravitational lensing and galaxy clustering provide compelling evidences that universe is an accelerating phase at present. They suggest that the content of universe is 70% dark energy, 25% dark matter and 5% ordinary matter [1]. In order to accommodate an accelerating universe within general relativity, it must be postulated a cosmic fluid with negative pressure. The simplest candidate implying this property is the cosmological constant. The others are quiessence, quintessence, brane-world models and Chaplygin gas.

In the simplest case, Chaplygin gas is a perfect fluid characterized by equation of state $p=-B/\rho$, where $B$ is a positive constant. Inspired by the fact that Chaplygin gas possesses a negative pressure it is used in cosmology to describe a transition from a decelerated cosmological expansion to the present epoch of a cosmic acceleration. Also it describes a unification of dark matter and dark energy. For a review about Chaplygin gas see [2].

In order to generalize Chaplygin Gas, one can take it of the form called by Modified Chaplygin Gas (MCG)

$$p = \gamma\rho - \frac{B}{\rho^{\alpha}} , \qquad 0 < \alpha \leq 1 , \quad 0 \leq \gamma \leq 1 \qquad (1)$$

We prefer to use MCG, because in this equation of state, radiation dominated phase of the early universe would be taken into account when $\gamma = 1/3$.

The paper is organized as follows: in section 2 we give brief information about Bianchi-type models and then write Einstein Field Equations (EFE) of all Bianchi types for a source with MCG. To get solutions of these equations, we use an approach proposed by M. K. Mak and T. Harko [3]. Then, we consider a special case, Bianchi I type model and then obtain solutions for cosmological parameters.

## 2. EFE FOR BIANCHI MODELS CHAPLYGIN GAS

Einstein Field Equations for 4-dimensional space-time can be written in the form

$$R_{ab} = T_{ab} - \frac{1}{2}g_{ab}T + \Lambda g_{ab} \qquad (2)$$

The general form of the metric for all Bianchi Models is

$$ds^2 = -dt^2 + \gamma_{ij}(t)\omega^i\omega^j \tag{3}$$

where $dt=\omega^0$ and $\omega^i$ : 1-forms which are independent of time. The metric can be simplified by taking diagonal matrices of the form, which is used whole models in class A and only V model in class B,

$$ds^2 = -dt^2 + R_1^2(t)(\omega^1)^2 + R_2^2(t)(\omega^2)^2 + R_3^2(t)(\omega^3)^2 \tag{4}$$

It is necessary for the other Bianchi models to get off-diagonal components of the metric. Here we are only interested in models which metrics are diagonizable. Then one can define a comoving volume of the universe, Hubble parameter and some relations of them in the following fashion: (i=1,2,3)

$$V(t) = R_1(t)R_2(t)R_3(t) \, , \, \frac{\dot{V}}{V} = H_1 + H_2 + H_3 \equiv 3H \, , \, H_i = \frac{\dot{R}_i}{R_i} \tag{5}$$

By using these definitions, EFE's and energy conservation equation for Bianchi models owning the metric (4) are

$$3\dot{H} + \sum_{i=1}^{3} H_i^2 = \Lambda - \frac{1}{2}\left[(3\gamma+1)\rho - \frac{3B}{\rho^\alpha}\right] \tag{6}$$

$$\frac{3\hat{a}_1}{R_1}(H_1 - H) = 0 \tag{7}$$

$$\frac{1}{V}\frac{d}{dt}(VH_i) = \Lambda + f_i(R_1, R_2, R_3) + \frac{1}{2}\left[(1-\gamma)\rho + \frac{B}{\rho^\alpha}\right] \tag{8}$$

$$\dot{\rho} + 3(\rho + p)H = 0 \tag{9}$$

Here $f_i$ functions are given in the [4] for all Bianchi-types respectively. Also $\hat{a}_1$ is a constant determining the class of models [5].

Substituting $p$ given in (1) into (9), we get the time evolution of the energy density of the MCG as

$$\rho = \left[\frac{B}{1+\gamma} + \frac{C}{V^{(1+\gamma)(1+\alpha)}}\right]^{\frac{1}{1+\alpha}} \tag{10}$$

By adding Eqs. (6) and defining $f = f_1 + f_2 + f_3$ we find

$$\frac{1}{V}\frac{d}{dt}(VH) = \frac{\ddot{V}}{3V}\Lambda + f(R_1, R_2, R_3) + \frac{1}{2}\left[(1-\gamma)\rho + \frac{B}{\rho^\alpha}\right] \tag{11}$$

Now to make calculations easier, we show rhs. of eqs. (6) with $\chi_i$ and eqs. (11) with $\chi$. Then from (11)

$$\dot{V} = \left\{\int 6V\chi dV + C_1\right\}^{1/2} = \phi^{1/2} \tag{12}$$

And the general solution of (11) is

$$t - t_0 = \int \frac{dV}{\phi^{1/2}} \tag{13}$$

Here we define $\Delta f_i = f_i - f$, so Hubble parameter, expansion scalar and anisotropy parameter become

$$H = 3\theta = \frac{\phi^{1/2}}{V} \tag{14}$$

$$A = \frac{1}{H^2 V^2}\left\{\frac{1}{27}\left(\int\frac{\Delta f_i}{H}dV\right)^2 + \frac{2}{9}\sum_i K_i \int\frac{\Delta f_i}{H}dV + \frac{1}{3}\sum_i K_i^2\right\} \tag{15}$$

In order to get exact solutions of EFE, we examine a special case which is Bianchi I model whose $f_i$ functions and $\hat{a}_1$ are equal to zero. Then substituting $f_i$, $\hat{a}_1$ and $\rho$ into upper equations we obtain the function $\chi$ for Bianchi I model. By using eqs.(18) and (19) we find how to relate of volume to time. From eqs. (23), the relations of time with expansion scalar and anisotropy parameter can be written by introducing two new variables $\tau$ and $v$ which we get from transformations $V=V_0 v$ and $\tau = \sqrt{3}B^{1/2(1+\alpha)}t$ with $V_0 =$ constant. It also needs some normalization of the arbitrary integration constants in the following way;

$$C\big/BV_0^{(1+\gamma)(1+\alpha)} = 1,\; C_1\big/3V_0^2 B^{1/(1+\alpha)} = 1,\; 6\Lambda\big/3B^{1/(1+\alpha)} = \lambda_0$$

## 3. DISCUSSIONS

By obtaining general solution of the field equations we have assumed non-zero cosmological constant $\Lambda$. Because both dark energy and dark matter have been taken into account by considering Chaplygin gas, we assume $\lambda_0=0$. Figure 1 shows how to evaluate expansion scalar with time for only $\alpha=1$. The difference of $\theta$ is not distinguishable for the other values of $\alpha$. The universe begins expanding at very high values and never stops. In Figure 2, the anisotropy of the universe is a non-zero value at the beginning of its evolution. Anisotropy decreases when the universe expands, but never goes to zero. Our results are consistent with those in reference [6].

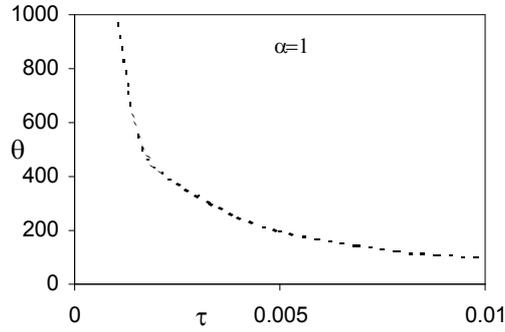

**FIGURE 1.** Expansion scalar $\theta$ of the Chaplygin gas filled Bianchi I universe as a function of dimensionless time parameter $\tau$ for $\alpha=1$

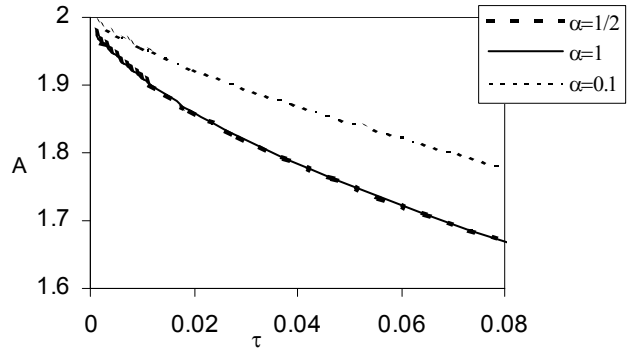

**FIGURE 2.** The anisotropy parameter A of the Chaplygin gas filled Bianchi I universe as a function of $\tau$ for different values of $\alpha$.

## REFERENCES

1. A.G. Riess et al., *Astron. J.* **116**, 1009 (1998); S. Perlmutter at al., *Astrophys. J.* **517**, 565 (1999) ; D.N. Spergel et al., *Astrophys. J. Supp.* **148**, 175 (2003) ; M. Tegmark et al., *Phys. Rev.D***69**, 103501 (2004)
2. V. Gorini et al. gr-qc/0403062 ; U. Alam et al. *Mon.Not.Roy.Astron.Soc.* 344, 1057, (2003)
3. M.K.Mak and T. Harko, *Phys.Rev. D***71**, 104022, (2005)
4. L.G. Jensen and J.A. Stein-Shabes, *Phys. Rev. D.*34 ,4, 1986 ; A. Harvey and D.Tsoubelis, *Phys. Rev. D.*15 , 10, 1977 ; T. Harko and M.K.Mak, *Class.Quant.Grav.* **21** 1489-1504, (2004).
5. J.Weinright and G.F.R. Ellis, *Dynamical Systems in Cosmology*, Cambridge University Press, Cambridge,    0 521 55457 8
6. B. Saha, *Chin.J.Phys*. 43, 1035-1043, (2005)